\documentclass[twocolumn,prb,showpacs,citeautoscript,floatfix]{revtex4}
\usepackage{graphicx}

\begin{document}

\title{Mott-Hubbard gap closure and structural phase transition in the oxyhalides TiOBr and TiOCl under pressure}

\author{C. A. Kuntscher}
\email[E-mail:~]{christine.kuntscher@physik.uni-augsburg.de}
\author{A. Pashkin}
\author{H. Hoffmann}
\author{S. Frank}
\author{M. Klemm}
\author{S. Horn}
\affiliation{Experimentalphysik 2, Universit\"at Augsburg, D-86135 Augsburg, Germany}
\author{A. Sch\"onleber}
\author{S. van Smaalen}
\affiliation{Laboratory of Crystallography, Universit\"at Bayreuth, 95440 D-Bayreuth, Germany}
\author{M. Hanfland}
\affiliation{European Synchrotron Radiation Facility, BP 220, F-38043 Grenoble, France}
\author{S. Glawion}
\affiliation{Experimentelle Physik 4, 
Universit\"at W\"urzburg, 97074 D-W\"urzburg, Germany}
\author{M. Sing}
\affiliation{Experimentelle Physik 4, 
Universit\"at W\"urzburg, 97074 D-W\"urzburg, Germany}
\author{R. Claessen}
\affiliation{Experimentelle Physik 4, 
Universit\"at W\"urzburg, 97074 D-W\"urzburg, Germany}
\date{\today}

\begin{abstract}
Pressure-dependent transmittance and reflectance spectra of TiOBr and TiOCl single crystals at room temperature 
suggest the closure of the Mott-Hubbard gap, i.e., the gap is filled with additional electronic states
extending down to the far-infrared range. 
According to pressure-dependent x-ray powder diffraction data the gap closure coincides with a structural 
phase transition. The transition in TiOBr
occurs at slightly lower pressure ($p$=14~GPa) compared to TiOCl ($p$=16~GPa) under hydrostatic conditions, 
which is discussed in terms of the chemical pressure effect. The results of pressure-dependent transmittance 
measurements on TiOBr at low temperatures reveal similar effects at 23~K, where the compound is in the 
spin-Peierls phase at ambient pressure.

\end{abstract}


\maketitle

\section{Introduction}
The layered compounds TiO$X$, where $X$=Br or Cl, are low-dimensional systems which show 
interesting magnetic and electronic properties.
Regarding the spin degree of freedom, at high temperature the system can be well described by a 
one-dimensional spin-1/2 nearest-neighbor Heisenberg model with a Bonner-Fisher type magnetic
susceptibility.\cite{Seidel03,Kataev03} Below the transition temperature T$_{c1}$, 
where T$_{c1}$=27~K for TiOBr and T$_{c1}$=67~K for TiOCl, TiO$X$ undergoes
a first-order phase transition to a spin-Peierls state with a dimerization of the chains of Ti 
atoms along the $b$ axis and a doubling of the unit cell.\cite{Seidel03,Caimi04,Shaz05}
Furthermore, an intermediate phase for the temperature range T$_{c1}$$<$T$<$T$_{c2}$ was found 
(with T$_{c2}$=47~K for TiOBr and T$_{c2}$=91~K for TiOCl), whose 
nature is now well established as an incommensurately modulated structure with a one-dimensional 
modulation in monoclinic symmetry.~\cite{Smaalen05}
Regarding the charge degree of freedom, the Ti ions have the electronic configuration $3d^1$. The 
$3d$ electrons are localized due to strong electronic correlations, and hence  
TiOBr and TiOCl are Mott-Hubbard insulators, with a charge gap of $\approx$2~eV. \cite{Ruckkamp05,Kuntscher06,Kuntscher07}
It was predicted that these materials exhibit a resonating valence bond state and high-temperature 
superconducti\-vi\-ty upon doping.\cite{Beynon93,Craco06}
However, up to now a metallization of TiO$X$ upon doping could not be achieved.\cite{Klemm08}

Recently it was shown that the optical response of both compounds changes drastically under pressure:
Above a critical pressure, the transmittance is suppressed and the reflectance increases 
in the infrared range. The changes could be attributed to additional electronic states filling 
the Mott-Hubbard gap and they suggest a closure of the gap at elevated pressures.\cite{Kuntscher06,Kuntscher07}
Under hydrostatic conditions the transition pressures are 14 and 16~GPa for TiOBr and TiOCl, 
respectively. Concurrent with the closure of the Mott-Hubbard gap a structural phase transition 
is observed.\cite{Kuntscher07}

This paper is a follow-up of the earlier, short publication\cite{Kuntscher07} and provides
details of the changes in the electronic properties and crystal structure of 
TiOBr and TiOCl induced by external pressure. 
The manuscript is organized as follows: After describing the experimental details in Sec.\
\ref{sectionexperiment}, we present in Sec.\ \ref{transmittance} the experimental results
obtained at room temperature, which suggest the closure of the Mott-Hubbard gap under pressure.
We also include low-temperature transmittance spectra of TiOBr at ambient and high pressure
in Sec.\ \ref{low-temperature results}. 
Sec.\ \ref{x-ray} focuses on the pressure-induced changes of the crystal structures for TiOBr and TiOCl.
In Sec.\ \ref{comparisontransitionpressures} we comment about a possible chemical pressure effect
in the system TiO$X$. In Sec.\ \ref{gap closure-structure} the relation between the closure of 
the Mott-Hubbard gap and the structural phase transition is discussed.  Finally, 
we summarize our results in Sec.\ \ref{summary}.

\begin{figure}[h]
\includegraphics[width=0.9\columnwidth]{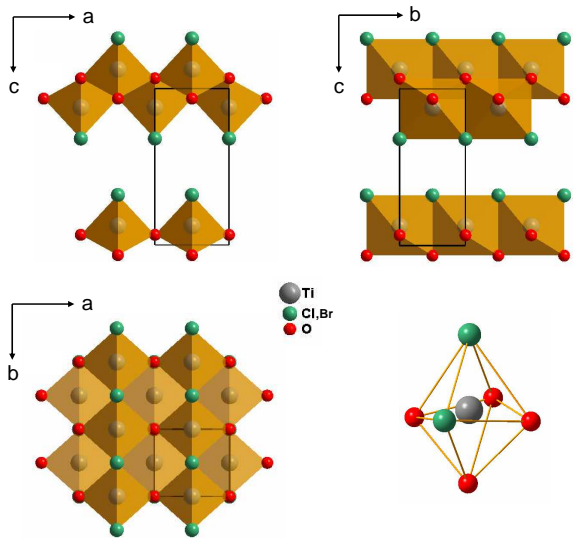}
\caption{Crystal structure of TiO$X$ ($X$=Br,Cl), viewed along
the $a$, $b$, and $c$ crystal axes, consisting of Ti-O 
bilayers parallel to the $ab$-plane and separated by layers of $X$ ions stacked along 
the $c$ direction.\cite{Schaefer58} The black lines mark the unit cell.
Also shown is the main building block of the crystal structure, namely the 
distorted TiO$_4$$X$$_2$ octahedron.}
\label{fig:crystalstructure}
\end{figure}

\section{Experiment}
\label{sectionexperiment} 

Single crystals of TiO$X$ ($X$=Br,Cl) were synthesized by chemical vapor transport 
technique. The TiO$X$ compounds crystallize in the space group $Pmmn$ at ambient conditions and consist of
distorted TiO$_4$$X_2$ octahedra.\cite{Schaefer58,Schnering72} The octahedra are arranged such that buckled Ti-O 
bilayers parallel to the $ab$-plane are formed, which are separated by layers of Br/Cl ions stacked along 
the $c$ direction. Fig.~\ref{fig:crystalstructure} shows the crystal structure viewed along 
the crystal axes $a$, $b$, and $c$. TiO$X$ crystals grow in the form of thin platelets with 
the surface parallel to the $ab$-plane. This is convenient for studies of the optical response
of the $ab$-plane.

In the pressure-dependent studies diamond anvil cells (DACs) were used for the generation 
of pressures. The applied pressures $p$ were determined with the ruby luminescence method.\cite{Mao86} 
For the transmittance measurements several pressure transmitting media were used; this leads
to small differences in the observed values of the critical pressure of phase 
transition, as expected.\cite{Frank06,Kuntscher06}
For the reflectance measurements finely ground CsI powder was chosen as pressure 
medium to insure direct contact of the sample with the diamond window.

Pressure-dependent transmittance and reflectance experi\-ments were
conducted at room temperature using a Bruker IFS 66v/S FT-IR
spectrometer with an infrared microscope (Bruker IRscope II).
For the generation of pressure we used a Syassen-Holzapfel DAC\cite{Huber77} equipped with 
type IIA diamonds suitable for infrared measurements.
Part of the measurements were carried out at the infrared beamline of the synchrotron radiation source ANKA,
where the same equipment is installed.
Further information on the pressure-dependent transmittance and reflectance measurements 
conducted at room temperature was included in the earlier publication.\cite{Kuntscher07}

For TiOBr the transmittance measurements under pressure were also conducted at
23~K for the frequency range 3100 - 15000~cm$^{-1}$ (0.38 - 1.9~eV). At 23~K TiOBr is 
in the spin-Peierls phase at ambient pressure. As pressure medium argon was used. 
The transmittance measurements on the sample in the DAC placed in the optical cryostat
(CryoVac KONTI cryostat) were performed using a home-built infrared microscope with
a large working distance.
This infrared microscope can be directly coupled to the FT-IR spectrometer and maintained
at the same pressure ($\approx$ 3 mbar), i.e., no window between the two devices is needed. 

Pressure-dependent x-ray powder diffraction measurements at room temperature 
were carried out at beamline ID09A of the European Synchrotron Radiation Facility at Grenoble.
Details about the experiments were described elsewhere (Ref.~\onlinecite{Kuntscher07}).

\begin{figure}[t]
\includegraphics[width=0.9\columnwidth]{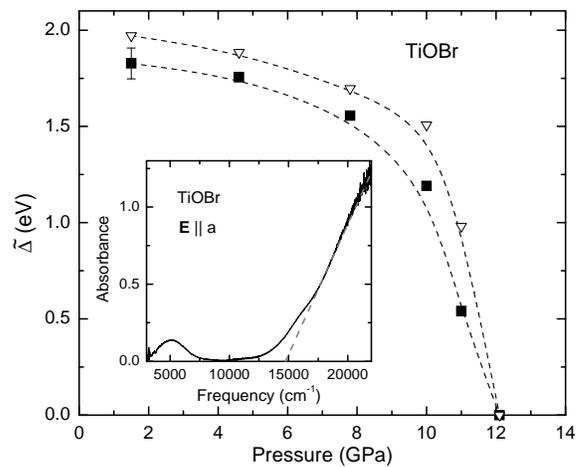}\\
\caption{Charge gap $\tilde{\Delta}$ (see text for definition) of TiOBr as a function of pressure
for {\bf E}$||$$a$ (full symbols) and {\bf E}$||$$b$ (open symbols) (pressure medium: CsI).
The dashed lines are guides to the eye.
Inset: Absorbance spectrum $A$($\omega$) of TiOBr for the lowest pressure (1.5~GPa), calculated 
according to $A$($\omega$)=log$_{10}$[1/$T$($\omega$)], together with the linear extrapolation of the 
absorption edge (dashed gray line) used to estimate $\tilde{\Delta}$.} \label{fig:gap}
\end{figure}

\section{Results and analysis}
\label{sectionresults}

\subsection{Pressure-dependent transmittance and reflectance at room temperature}
\label{transmittance}

Pressure-dependent transmittance measurements on TiOBr and TiOCl were carried out for
several pressure transmitting media. In Refs.~\onlinecite{Kuntscher06,Kuntscher07} we already
showed the spectra of TiOBr and TiOCl for argon and CsI as pressure media, respectively. 
The transmittance spectra reveal the characteristic excitations in the materials, namely
the electronic transitions between the lower and upper Hubbard gap, resulting in a strong
suppression of the transmittance above $\approx$2~eV. Furthermore, absorptions occur due to 
excitations across the crystal-field split Ti$3d$ energy levels (called orbital 
excitations in the following) located for TiOBr (TiOCl) at 0.63~eV (0.66~eV) for {\bf E}$||$$a$
and at 1.35~eV (1.53~eV)  for {\bf E}$||$$b$ at ambient conditions.

First, one notices that in TiOBr the orbital excitations are slightly redshifted compared to TiOCl. 
This can be explained by the chemical pressure effect in the system TiO$X$: 
Based on the g tensors measured by electron spin resonance\cite{Kato05} the crystal 
field splittings in TiOBr and Ti\nolinebreak OCl were obtained. The smaller crystal field splitting in TiOBr 
could be attributed to the larger size of the Br$^{-}$ ion compared to the Cl$^{-}$ ion, causing 
a larger volume of the TiO$_4$$X$$_2$ octahedra (see Fig.~\ref{fig:crystalstructure}) and hence 
a weaker crystal field.\cite{Kato05}

\begin{figure}[t]
\includegraphics[width=1\columnwidth]{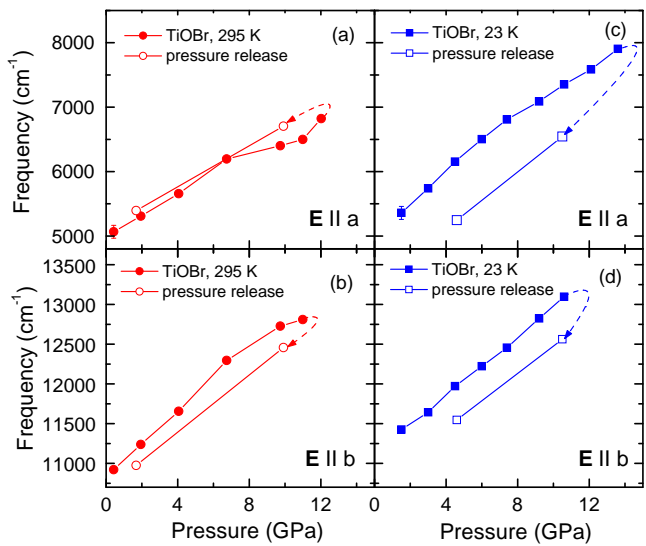}\\
\caption{(Color online) Frequency of the orbital excitations in TiOBr as a function of pressure: 
at room temperature
for the polarization (a) {\bf E}$||$$a$ and (b) {\bf E}$||$$b$; at 23~K for the 
polarization (c) {\bf E}$||$$a$ and (d) {\bf E}$||$$b$ (pressure medium: argon). 
The full symbols denote the results with increasing pressure; open symbols
denote the results upon pressure release. Lines are guides to the eye.
} \label{fig:orbital}
\end{figure}

%\begin{figure}[t]
%\includegraphics[width=0.8\columnwidth]{refl.eps}
%\caption{(Color online) Room-temperature reflectance $R_{\rm s-d}$ of TiOBr as a function of pressure for 
%the polarization (a) {\bf E}$||$$a$ and (b) {\bf E}$||$$b$ (pressure medium: CsI).\cite{Kuntscher07} 
%Arrows illustrate the changes with increasing pressure.}
%\label{fig:reflectivity}
%\end{figure}

With increasing pressure one observes the following changes for both compounds
along the two studied polarization directions:
(i) a blueshift of the orbital excitations; 
(ii) the absorption edge due to excitations across the charge gap shifts to smaller energies 
with increasing pressure, and above 11~GPa (12~GPa) the overall transmittance is strongly suppressed 
in TiOBr (TiOCl). These results were obtained with CsI as pressure medium;
when a more hydrostatic pressure medium is used (see Table~\ref{tab:comparison} and the results
in Ref.~ \onlinecite{Kuntscher07}), the suppression of the transmittance occurs at somewhat 
higher pressure ($\Delta$$p$$\approx$4~GPa).

We estimated the charge gap, $\tilde{\Delta}$, by a linear extrapolation of the steep absorption 
edge. This is illustrated in the inset of Fig.~\ref{fig:gap}, where we show the absorbance spectrum 
$A$($\omega$) of TiOBr for the lowest pressure (1.5~GPa), calculated from the transmittance $T$($\omega$) 
according to $A$($\omega$)=log$_{10}$[1/$T$($\omega$)], together with the linear extrapolation of the 
absorption edge. The intersection of the linear extrapolation with the horizontal axis
was taken as an estimate of the charge gap. 
Starting from the lowest applied pressure, $\tilde{\Delta}$ initially slightly decreases
with increasing pressure, and above $\approx$10~GPa it rapidly drops to zero (Fig.~\ref{fig:gap}). 
%This finding indicates the closure of the Mott-Hubbard gap, i.e., a transition from an insulating to a metallic state.
%We ascribe the gap closure to additional, pressure-induced excitations in the infrared range,
%as discussed in more detail below. Concomitantly with the shift of the absorption edge
%occurs a change of the sample color from red to black, i.e., the sample is opaque
%at high pressure (see Fig.~\ref{fig:picture}). 
Similar observations were made earlier for TiOCl,\cite{Kuntscher06} with the onset of rapid decrease
of $\tilde{\Delta}$ at $p$$\approx$12~GPa.
% and with a complete closure of the charge gap at $p$$\approx$14~GPa.
%Fig.~\ref{fig:picture} also provides a view of the sample at 2.3~GPa during pressure release.
%Obviously, the pressure-induced change of the sample color is not completely reversed, since 
%the sample remains partly black. A similar observation was made earlier for
%TiOCl.\cite{Kuntscher06}

The pressure dependence of the frequencies of the orbital excitations in TiOBr were obtained
by fitting the absorption features in the transmittance spectra with Gaussian functions.
The results are depicted in Fig.~\ref{fig:orbital}. With increasing pressure the orbital
excitations shift to higher frequencies in a linear fashion. This shift could be attributed
to a monotonically increasing strength of the crystal field related to the decreasing volume
of the TiO$_4$Br$_2$ octahedra. External pressure could also induce a change in the octahedral distortion
and related alterations of the crystal field. 
One furthermore notices a small difference in the frequency of the orbital 
excitations for pressure increase and decrease observed in the direction {\bf E}$||$$b$ 
[Fig.~\ref{fig:orbital}(b)], which suggests that the pressure-induced octahedral volume
decrease and/or octahedral distortion are not completely reversible. This non-reversibility 
is more obvious at low temperatures and will be discussed in Sec. \ref{low-temperature results}.

\begin{figure}[t]
\includegraphics[width=1\columnwidth]{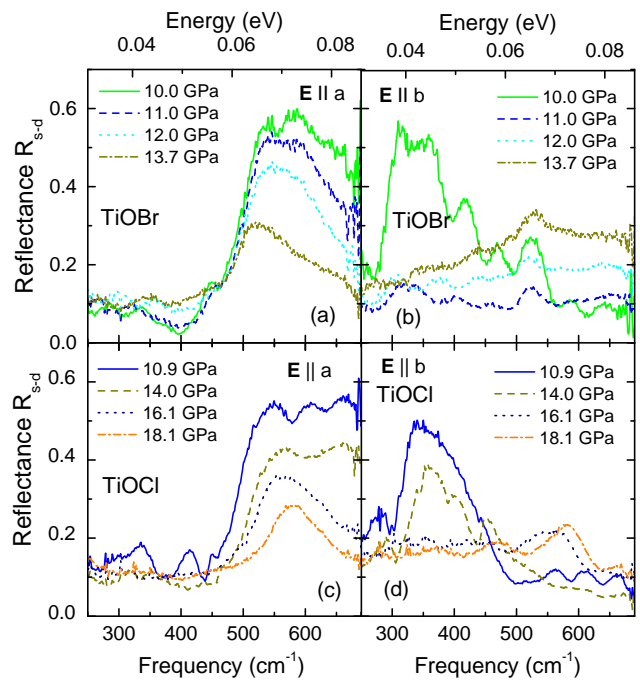}
\caption{(Color online) Far-infrared reflectance $R_{\rm s-d}$ of TiOBr and TiOCl at room temperature 
as a function of pressure, for the polarization {\bf E}$||$$a$ [(a) and (c), resp.] and 
{\bf E}$||$$b$ [(b) and (d), resp.] (pressure medium: CsI).}
\label{fig:FIR-reflectivity}
\end{figure}

%\begin{figure}[t]
%\includegraphics[width=1\columnwidth]{cond-Br.eps}
%\caption{(Color online) Reflectance spectrum $R_{\rm s-d}$ of TiOBr at 14~GPa for the polarization 
%(a) {\bf E}$||$$a$ and (b) {\bf E}$||$$b$ together with the Drude-Lorentz fits (dashed lines)  
%(pressure medium: CsI). Based on the Drude-Lorentz fits, the real part of the optical conductivity, 
%$\sigma_1$($\omega$), for the polarization (c) {\bf E}$||$$a$ and (d) {\bf E}$||$$b$ was obtained.}
%\label{fig:cond}
%\end{figure}

\begin{figure}[t]
\includegraphics[width=0.85\columnwidth]{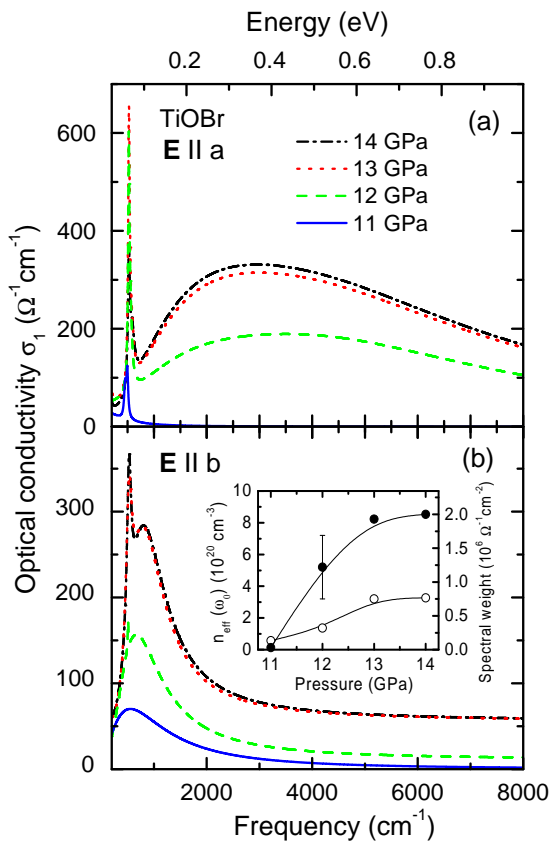}
\caption{(Color online) Real part of the optical conductivity of TiOBr as a function of pressure 
for the polarization (a) {\bf E}$||$$a$ and (b) {\bf E}$||$$b$ obtained by Drude-Lorentz fitting 
of the pressure-dependent reflectance data $R_{\rm s-d}$. Inset: Total effective carrier density, $n_{eff}$,
and spectral weight calculated by integrating the real part of the optical conductivity (see text) up to 
$\omega_0$=8000~cm$^{-1}$ for {\bf E}$||$$a$ (filled circles) and {\bf E}$||$$b$ (open circles).
Lines are guides to the eye.}
\label{fig:cond-all}
\end{figure}

Additional information about the pressure-induced changes in the optical response 
were obtained by reflectance measurements on TiOBr and TiOCl at high pressures. 
%Since the results for both compounds are very similar and have already been published in part for TiOCl,
%we will mainly focus on TiOBr in the following. Fig.~\ref{fig:reflectivity} shows the polarization-dependent 
%reflectance $R_{\rm s-d}$ of TiOBr in a broad frequency range for pressures $p$$\geq$10~GPa. 
%This is the pressure range, where TiOBr becomes opaque (see Fig.~\ref{fig:transmittance}); 
%reflectance data below this pressure range cannot be analyzed quantitatively.
%Between 10 and 11~GPa the reflectance spectrum changes fundamentally, in particular in the 
%frequency range below 700~cm$^{-1}$. With further pressure increase ($p$$\geq$11~GPa)
%the reflectance of TiOBr increases in the whole studied frequency range (far-infrared to 
%near-infrared) and saturates above 13~GPa. 
The most drastic changes occur in the far-infrared range, as illustrated for both compounds
in Figs.~\ref{fig:FIR-reflectivity} (a) and (b)
(for pressure-dependent reflectance spectra over a broader frequency range, 
see Refs.~\onlinecite{Kuntscher06,Kuntscher07}).
%In Fig.~\ref{fig:FIR-reflectivity} (a) and (b) we present the pressure-dependent reflectance spectra 
%$R_{\rm s-d}$ of TiOBr in the far-infrared region, where the most pronounced changes occur. 
In case of TiOBr, the shape of the spectrum changes drastically for {\bf E}$||$$b$:
At 10~GPa the spectrum consists of a peak-like feature between 300 and 450~cm$^{-1}$, whereas
for pressures $\geq$11~GPa it is almost flat with a peak at around 520~cm$^{-1}$.
The pressure-induced changes in the far-infrared reflectance spectra
$R_{\rm s-d}$ of TiOCl are very similar to those of TiOBr [see 
Figs.~\ref{fig:FIR-reflectivity} (c) and (d)]. However, for TiOCl the changes occur at somewhat higher
pressure, as discussed in Sec. \ref{comparisontransitionpressures}.
For higher frequencies the overall reflectance increases for both compounds and saturates.\cite{Kuntscher06,Kuntscher07}

%\subsection{Pressure-induced metallization at room temperature}
%\label{metallization}

The suppression of the transmittance in TiO$X$ at high pressures suggests the occurrence  of
new excitations in the infrared frequency range.
More information about these additional excitations were obtained by fitting the 
high-pressure ($p$$>$10~GPa) reflectance spectra $R_{\rm s-d}$ with the Drude-Lorentz model combined with the
normal-incidence Fresnel equation, taking into account the diamond-sample interface:
\begin{equation}
R_{s-d} =\left| \frac{n_{\rm dia}-\sqrt{\epsilon_s}}{n_{\rm
dia}+\sqrt{\epsilon_s}}\right|^2 , \epsilon_s = \epsilon_\infty +
\frac{i \sigma}{\epsilon_0 \omega} \quad ,
\end{equation}
where $\epsilon_s$ is the complex dielectric function of the
sample and $\epsilon_\infty$ is the background dielectric constant (here $\epsilon_\infty$$\approx$3). 
From the function $\epsilon_s$($\omega$) the real part of the 
optical conductivity, $\sigma_1$($\omega$), can be calculated.
Notice that only reflectance data above 10~GPa can be analyzed quantitatively because of 
the partial transparency of the sample below this critical pressure.
%As an example, we present in Fig.~\ref{fig:cond}
%the reflectance spectrum $R_{\rm s-d}$ of TiOBr at the highest studied pressure (14~GPa) for 
%the polarization (a) {\bf E}$||$$a$ and (b) {\bf E}$||$$b$ together with the Drude-Lorentz 
%fits (dashed lines). The two lower graphs [Fig.~\ref{fig:cond} (c) and (d)] show the 
%polarization-dependent real part of the optical conductivity, $\sigma_1$($\omega$), obtained 
%from the fitted model.
%Accordingly, we find additional excitations in the infrared frequency range, extending 
%down to the far-infrared. This evidences the metallization of the sample at high pressures. 
%The additional excitations include broad excitations, which cannot
%be attributed to phonon excitations, in contrast to the optical conductivity spectrum
%in the insulating phase.\cite{Caimi04a} We obtain similar results for the analog compound TiOCl (see
%Ref. \onlinecite{Kuntscher06}).

The evolution of the optical conductivity of TiOBr with pressure is shown in 
Fig.~\ref{fig:cond-all}. We find additional excitations in the infrared range, 
extending down to the far-infrared. These additional excitations include broad excitations, 
which cannot be attributed to phonon excitations, in contrast to the optical conductivity 
spectrum in the insulating phase.\cite{Caimi04a}
Thus, the Mott-Hubbard gap is gradually filled with additional electronic states down
to at least 200~cm$^{-1}$ (24~meV). This finding suggests the closure of the Mott-Hubbard gap
above $p$=10~GPa. 

With increasing pressure the spectral weight of the pressure-induced features increases,
with a saturation setting in at around 13~GPa. 
From the spectral weight analysis one can extract the effective density of carriers, $n_{eff}$,
involved in the excitations up to $\omega_0$ according to 
\begin{equation} \label{carrierdensity}
n_{eff}(\omega_0)=(2m_0 / \pi e^2)\int_0^{\omega_0}\sigma_1(\omega)d\omega \quad ,
\end{equation}
with the free electron mass, $m_0$.
In the inset of Fig.~\ref{fig:cond-all}(b) $n_{eff}$($\omega_0$=8000~cm$^{-1}$)
is plotted as a function of pressure $p$. $n_{eff}(p)$ illustrates the saturation of
the spectral features at high pressures.

Also for TiOCl the spectral weight of the pressure-induced excitations increases
with increasing pressure and saturates, as presented in the earlier publication.\cite{Kuntscher06} 
The saturation, however, happens at somewhat higher pressure ($\approx$15~GPa) compared 
to TiOBr. In fact, all the pressure-induced effects occur in TiOCl at slightly higher 
pressures ($\Delta$$p$$\approx$2~GPa) compared to TiOBr. This pressure difference will be 
discussed in more detail in Sec.~\ref{comparisontransitionpressures}.

\begin{figure}[t]
\includegraphics[width=0.8\columnwidth]{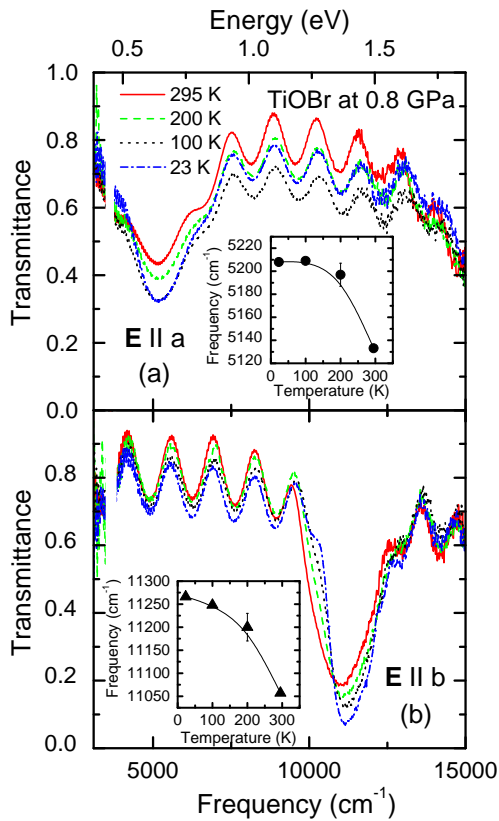}
\caption{(Color online) Transmittance $T$($\omega$)=I$_{s}$($\omega$)/I$_{r}$($\omega$)
(see text for definitions) of TiOBr as a function of temperature for the lowest pressure (0.8~GPa), 
for the polarization (a) {\bf E}$||$$a$ and (b) {\bf E}$||$$b$ (pressure medium: argon). 
Insets: Frequency of orbital excitations as a function of temperature. Lines are guides to the eye.}
\label{fig:transmittance-T}
\end{figure}

\begin{figure}[t]
\includegraphics[width=0.8\columnwidth]{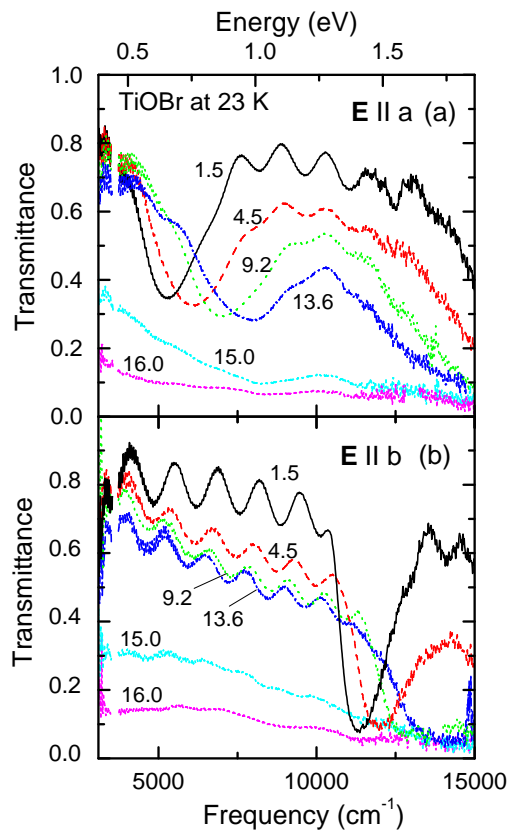}
\caption{(Color online) Transmittance $T$($\omega$)=I$_{s}$($\omega$)/I$_{r}$($\omega$)
(see text for definitions) of TiOBr as a function of pressure at 23~K, 
for the polarization (a) {\bf E}$||$$a$ and (b) {\bf E}$||$$b$ (pressure medium: argon).}
\label{fig:transmittance-P}
\end{figure}

\subsection{Pressure-dependent transmittance of TiOBr at low temperatures}
\label{low-temperature results}

We have furthermore checked the stability of the insulating spin-Peierls phase of TiOBr
by pressure-dependent transmittance measurements at low temperatures in the near-infrared 
frequency range (3100 - 15000~cm$^{-1}$). As mentioned in the introduction, TiOBr
undergoes two phase transitions as a function of temperature: Upon temperature increase, 
a first order transition takes place at $T_{c1}$=27~K from the spin-Peierls ground state into 
an intermediate phase with an incommensurate superstructure.\cite{Smaalen05} An additional, second-order 
phase transition is found at $T_{c2}$=47~K, where the material changes from the intermediate 
phase to the one-dimensional antiferromagnetic phase at high temperature. 

Starting from room temperature and low-pressure (0.8~GPa) conditions, transmittance measurements on 
TiOBr were carried out upon temperature decrease. Fig.~\ref{fig:transmittance-T} shows the temperature-dependent
transmittance spectra for the polarizations {\bf E}$||$$a$ and {\bf E}$||$$b$. The oscillations
in the spectra are Fabry-Perot resonances due to multiple reflections within the thin
sample platelet. With decreasing temperature one notices
a small but significant shift of the orbital excitations to higher
frequencies, as illustrated in the insets of Fig.~\ref{fig:transmittance-T}.
The most pronounced changes occur between 295 and 200~K and can be attributed to the thermal
contraction of the lattice while cooling down, leading to a smaller volume of the TiO$_4$Br$_2$
octahedra and thus to a stronger crystal field.\cite{Kato05}
Below 100~K the orbital excitations hardly shift with temperature, which suggests
that the structural changes occuring at the phase transitions at $T_{c1}$=27~K and $T_{c2}$=47~K 
have only a small effect on the TiO$_4$Br$_2$ octahedra and hence on the crystal field.
This in agreement with an earlier work showing that the orbital degree of freedom is
irrelevant for the low-energy physics,  in particular the exotic spin-Peierls behavior
with two successive phase transitions.\cite{Ruckkamp05} 

%\begin{figure}[t]
%\includegraphics[width=0.8\columnwidth]{x-rayBr.eps}
%\caption{(Color online) Room-temperature x-ray powder diffraction diagrams of TiOBr 
%at high pressures ($\lambda$=0.4128~\AA) together with the LeBail fits (pressure medium: helium). 
%For the lowest applied pressure (1.7~GPa) the difference curve ($I_{obs}-I_{calc}$) between 
%the diffraction diagram and the LeBail fit is shown. 
%Markers show the calculated peak positions for the ambient-pressure phase. Above 14~GPa the 
%diffraction diagram can no longer be described by the ambient-pressure crystal symmetry.
%Arrows indicate the diffraction peaks with the most obvious discrepancy between the 
%data and the LeBail fitting curve.}
%\label{fig:x-ray-TiOBr}
%\end{figure}

\begin{figure}[t]
\includegraphics[width=0.8\columnwidth]{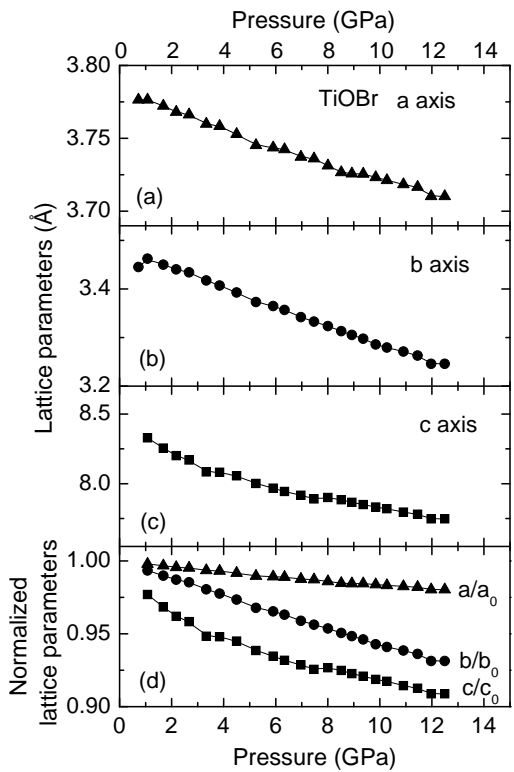}
\caption{(Color online) Lattice parameters of TiOBr at room temperature as a function of pressure
(pressure medium: helium).
(a)-(c) Lattice parameters $a$, $b$, $c$. 
(d) Lattice parameters $a$, $b$, $c$ normalized to 
their respective zero-pressure values. Lines are guides to the eye.}
\label{fig:lattice-TiOBr}
\end{figure}

At 23~K, where the sample is in the spin-Peierls state for ambient pressure, 
transmittance spectra were recorded for several pressures (see Fig.~\ref{fig:transmittance-P}). 
Similar to the room-temperature results, the transmittance is suppressed over the whole studied
frequency range above a certain pressure; however, at 23~K the suppression occurs only above 
$\approx$16~GPa, compared to the room-temperature transition pressure of 14~GPa.

We also followed the pressure-induced shifts of the orbital excitations at 23~K [Figs.~\ref{fig:orbital} (c) 
and (d)]. With increasing pressure the frequencies of the 
orbital excitations increase linearly with increasing pressure. Like for the room-temperature
results, we relate this shift to a pressure-induced decrease of the octahedral volume and a possible
change in octahedral distortion, causing a change in the crystal field (see Sec.\ \ref{transmittance}).
At room-temperature we noticed a small difference in the frequencies of the orbital 
excitations for pressure increase and decrease. This difference is much more pronounced at 23~K.
For example, for {\bf E}$||$$a$ already at around 4~GPa during pressure release the ambient-pressure 
excitation energy of $\approx$5370~cm$^{-1}$, and thus the ambient-pressure crystal field
strength, has been reached [see Fig.~\ref{fig:orbital}(c)].

\begin{figure}[t]
\includegraphics[width=0.8\columnwidth]{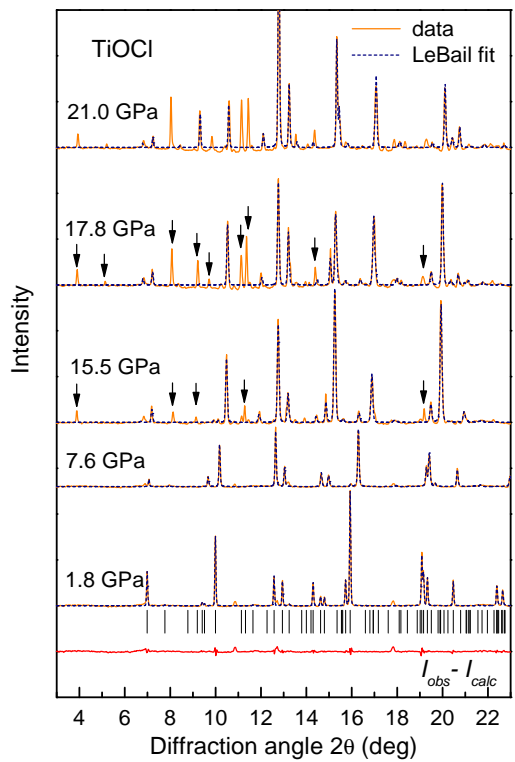}
\caption{(Color online) Room-temperature x-ray powder diffraction diagrams of TiOCl at high pressures 
($\lambda$= 0.4128~\AA) together with the LeBail fits (pressure medium: helium). 
For the lowest applied pressure (1.8~GPa) the 
difference curve ($I_{obs}-I_{calc}$) between the diffraction diagram and the LeBail fit is shown. 
Markers show the calculated peak positions for the ambient-pressure phase. Above 15.5~GPa the 
diffraction diagram can no longer be described by the ambient-pressure crystal symmetry.
Arrows indicate the diffraction peaks with the most obvious discrepancy between the 
data and the LeBail fitting curve.}
\label{fig:x-ray-TiOCl}
\end{figure}

\begin{figure}[t]
\includegraphics[width=0.8\columnwidth]{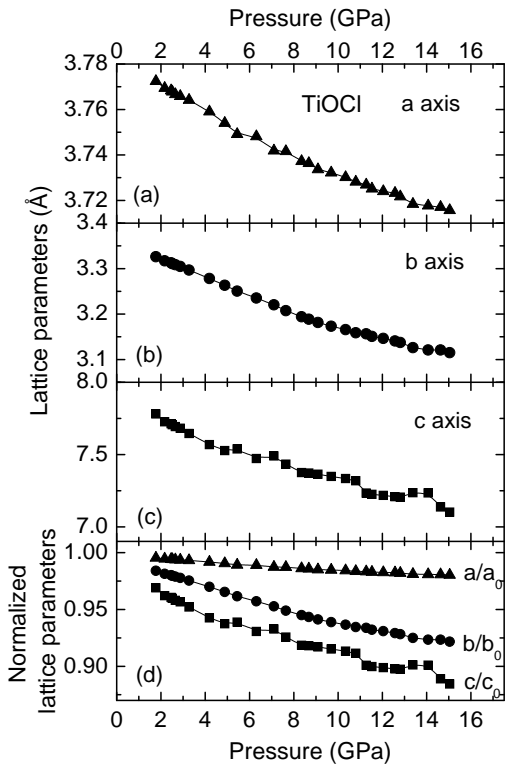}
\caption{(Color online) Lattice parameters of TiOCl at room temperature as a function of pressure
(pressure medium: helium).
(a)-(c) Lattice parameters $a$, $b$, $c$. 
(d) Lattice parameters $a$, $b$, $c$ normalized to 
their respective zero-pressure values. Lines are guides to the eye.}
\label{fig:lattice-TiOCl}
\end{figure}

\subsection{Pressure-induced structural phase transition at room temperature}
\label{x-ray}

For the understanding of the drastic changes in the optical response under pressure, we carried
out x-ray powder diffraction measurements on TiOBr and TiOCl at room temperature as a function of pressure. 
A typical diffraction pattern (not shown) does not consist of concentric rings as expected for powder diffraction 
data, but it contains separate spots. This is due to the fact that it was not possible to 
produce good TiO$X$ powders with homogenous grain size distributions and random orientations
because of the platelet-like habits and the softnesses of the crystallites. Instead, the 
crystallites inside the DAC orient their $c$ crystal 
axis preferentially perpendicular to the diamond anvil surface, i.e., along the direction of 
incidence of the x-radiation. Therefore, Rietveld refinements of the diffraction patterns
could not be carried out. Nevertheless LeBail fits of the diffraction patterns could be 
accomplished, in order to determine the unit cell volume and the lattice parameters as a function of pressure.

The room-temperature diffraction diagrams of TiOBr can be well fitted with the ambient-pressure 
crystal structure (space group $Pmmn$) at low pressures, as demonstrated in Ref.~\onlinecite{Kuntscher07}.
The lattice parameters of TiOBr as a function of pressure, as obtained from the LeBail fitting, are 
presented in Fig.~\ref{fig:lattice-TiOBr}. The changes of the lattice parameters $a$ and $b$ 
with pressure are linear over a large pressure range. The behavior of the lattice 
parameters $c$ rather follows a sublinear fashion.
In Fig.~\ref{fig:lattice-TiOBr} 
we also show the lattice parameters $a$, $b$, $c$ normalized to their respective zero-pressure values
as a function of pressure [Fig.~\ref{fig:lattice-TiOBr} (d)]. According to these results, TiOBr has 
a very anisotropic compressibility, with the largest compressibility along the $c$ axis, i.e., 
the stacking axis of the buckled Ti-O bilayers.

At around 14~GPa the diffraction diagram of TiOBr undergoes pronounced changes
and is no longer compatible with the ambient-pressure crystal structure symmetry.\cite{Kuntscher07} 
We can therefore conclude that TiOBr undergoes a structural phase transition at 14~GPa.

We also include the corresponding results from the pressure-dependent x-ray powder diffraction 
on TiOCl, namely the room-temperature diffraction diagrams for selected pressures together with the LeBail 
fits (Fig.~\ref{fig:x-ray-TiOCl}) as well as the lattice parameters as a function of pressure
extracted by the LeBail fits (Fig.~\ref{fig:lattice-TiOCl}). 
The sublinear dependence on pressure is obvious for all three lattice parameters.
Pronounced changes of the diffraction diagram occur at 15.5~GPa indicating a 
pressure-induced structural phase transition in TiOCl, similar like in TiOBr. 
For both compounds the pressure-induced changes are reversible in terms of the positions
of the diffraction peaks.

\begin{figure}[t]
\includegraphics[width=0.85\columnwidth]{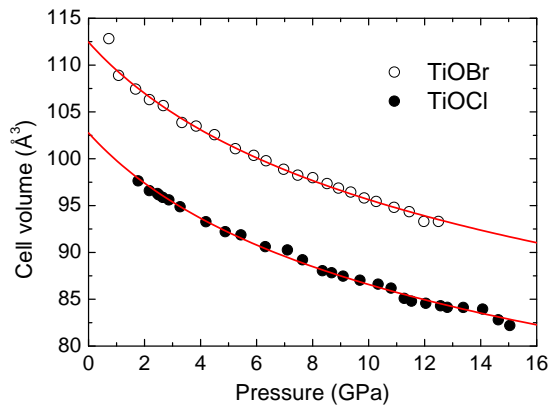}
\caption{(Color online) Unit cell volume $V$ of TiOCl and TiOBr as a function of pressure $P$. 
The full, red (gray) lines are fits according to Eq.~(\ref{Murnaghan}).
}
\label{fig:volume}
\end{figure}

From the lattice parameters the pressure dependence of the unit cell volume $V$ for both compounds
was obtained. In Fig.~\ref{fig:volume} we plot $V(p)$ together with a fit according to the Murnaghan 
equation \cite{Murnaghan44}
\begin{equation}
\label{Murnaghan}
V(p) = V_0 [(B'/B_0)p+1]^{-1/B'}
\end{equation}
with the bulk modulus $B_0=-dp/dlnV$ and its derivative $B'$ at zero pressure. The ambient-pressure
unit cell volume $V_0$ was kept fixed at the experimental value of 112.4(5)~\AA$^3$ [102.7(6)~\AA$^3$] 
for TiOBr (TiOCl).~\cite{Sasaki05} The bulk moduli $B_0$ evaluated according to the Murnaghan equation
are 33.7$\pm$ 0.8~GPa and 31.0$\pm$ 0.9~GPa for TiOBr and TiOCl, and the derivatives $B'$ are 
6.9$\pm$ 0.3 and 6.7$\pm$ 0.3, respectively. 
The bulk modulus of TiOBr is 
slightly larger than that of TiOCl, i.e., TiOBr is slightly less compressible than TiOCl.
Furthermore, one notices that the pressure derivative $B'$ of both compounds is significantly
larger compared to the value $B'$$\approx$4 typically found for three-dimensional solids with
isotropic elastic properties. The enhanced value of $B'$ thus suggests anisotropic compression properties
of TiO$X$. It is interesting to note that the bulk modulus $B_0$ and its derivative $B'$
of TiO$X$ are close to the corresponding values found for graphite ($B_0$=33.8~GPa, $B'$=8.9).\cite{Hanfland89}

\begin{table*}
\caption{\label{tab:comparison} Comparison of transition pressures of TiOCl and TiOBr at room temperature 
obtained from transmittance, reflectance, and x-ray powder diffraction measurements for different
pressure transmitting media.}
\begin{ruledtabular}
\begin{tabular}{cccccc}
material       & transmittance  & transmittance & transmittance & reflectance & x-ray diffraction \\
       & (CsI)           &  (argon) & (alcohol mixture)  & (CsI) & (helium) \\
\hline
TiOCl & 12~GPa    &   16~GPa         & $\approx$16~GPa & 12~GPa & 15.5~GPa  \\
\hline
TiOBr & 10-11~GPa &   14~GPa (295~K) & not measured    & 10-11~GPa & 14~GPa \\
      &           &   16~GPa (23~K)
\end{tabular}
\end{ruledtabular}
\end{table*}

\section{Discussion}
\label{sectiondiscussion}

\subsection{Comparison of transition pressures: TiOBr and TiOCl}
\label{comparisontransitionpressures}

In Table \ref{tab:comparison} we compare the transition pressures of TiOBr and TiOCl at room 
temperature obtained by different experimental techniques (transmittance, reflectance, x-ray
powder diffraction) and for different pressure media. 
First, comparing the corresponding results for the two compounds, one notices a pressure difference
of $\approx$2~GPa. This suggests the existence of some sort of chemical pressure effect in TiO$X$.

A starting point for the understanding of this finding could be a comparison of the ambient-pressure 
lattice parameters. The lattice parameters $b$ and $c$ of TiOBr ($a$=3.785~\AA, $b$=3.485~\AA, 
$c$=8.525~\AA) are significantly
larger than those of TiOCl ($a$=3.789~\AA, $b$=3.365~\AA, $c$=8.060~\AA).~\cite{Sasaki05,Kataev03}
The difference is most pronounced for the $c$ axis; here, the larger value in TiOBr can be 
attributed to the larger size of the Br$^-$ ions, which form layers separating the 
buckled Ti-O bilayers. 
Naively, one would then expect a {\it higher} pressure to induce the transition in TiOBr
compared to TiOCl, which is in contradiction to our findings. 
Thus, not the distance between the Ti-O bilayers but the pressure-induced crystal
structure changes {\it within} the bilayers seem to be the crucial parameter for inducing the 
closure of the Mott-Hubbard gap in TiO$X$. This furthermore suggests that the high-pressure phase 
has a dimensionality of less than three, being mainly confined to the buckled 
Ti-O bilayers.

\begin{figure}[t]
\includegraphics[width=0.85\columnwidth]{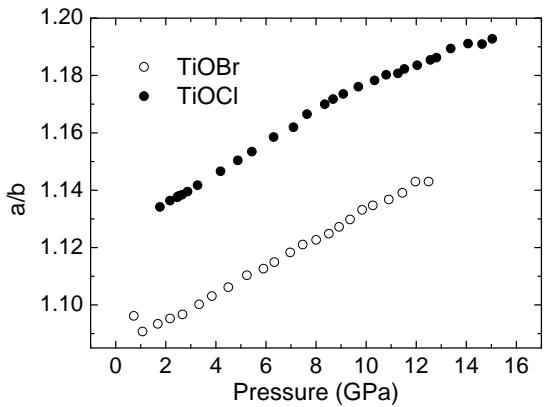}
\caption{Pressure-dependent ratio of the lattice parameters $a$ and $b$ for the TiOBr and TiOCl
at room temperature. The ratio $a$/$b$ as a function of pressure follows a linear behavior.}
\label{fig:ratio}
\end{figure}

A two-dimensional character of the high-pressure phase could indeed explain the difference in the
critical pressures for TiOBr and TiOCl: At ambient conditions the one-dimensional character of 
TiOBr is weaker than in TiOCl, since the lattice parameter ratio ($a$/$b$) in TiOBr ($a$/$b$=1.086) 
is smaller than in TiOCl ($a$/$b$=1.126).\cite{Sasaki05}
This is consistent with magnetic susceptibility measurements showing a larger deviation
from the Bonner-Fisher type behavior above the spin-Peierls transition for TiOBr compared
to TiOCl.\cite{Kato05}  The more two-dimensional character of TiOBr was also demonstrated
by recent photoemission experiments and supported by density-functional calculations.\cite{Hoinkis07}
Hence, in the case of TiOBr less pressure would be needed to drive the system into a (prospective) 
two-dimensional, high-pressure phase. 

Our pressure-dependent crystal structure data can provide a test of this picture (two-dimensional
character of the high-pressure phase):
In Fig.~\ref{fig:ratio} we plot the ratio $a/b$ for both studied compounds as a function of pressure.
This ratio should decrease towards the value 1 under pressure in the case of a two-dimensional high-pressure 
phase.
Instead, the ratio $a/b$ {\it increases} with increasing pressure for both compounds:
TiOBr and TiOCl become more one-dimensional under pressure.
Obviously, a different criterium regarding the changes of the crystal structure with applied pressure 
has to be used, in order to explain the difference of $\approx$2~GPa for the critical pressures
of TiOBr and TiOCl. At this point we can only speculate about possible criteria for the
Mott-Hubbard gap closure -- like a critical Ti-Ti distance along $b$ or $a$ direction -- since information about 
the shifts of the atomic coordinates under pressure is not available.

\subsection{Mott-Hubbard gap closure and structural phase transition}
\label{gap closure-structure}

Under hydrostatic conditions, i.e., for argon as pressure medium, the 
closure of the Mott-Hubbard gap in TiOBr (TiOCl) occurs at 14~GPa (16~GPa) 
(Table \ref{tab:comparison}). Under less hydrostatic conditions, i.e., using CsI as pressure
medium, the gap closure happens at somewhat lower pressure ($\Delta$$P$$\approx$4~GPa).
This offset in the transition pressure for different types of pressure media has been reported
earlier for TiOCl and LaTiO$_{3.41}$.\cite{Frank06,Kuntscher06}
The important finding is that under similar hydrostatic conditions the closure of
the Mott-Hubbard gap in TiO$X$ coincides with a structural phase transition, as 
demonstrated by our pressure-dependent x-ray powder diffraction data (Table \ref{tab:comparison}). 
Therefore, the gap closure in TiO$X$ is not of purely electronical origin, but the lattice 
degree of freedom has to be taken into account too.

In this regard it is interesting to compare the results for TiOBr and TiOCl with typical examples of 
bandwidth-controlled Mott transitions under external pressure, which are discussed in literature.
One finds the general observation that the Mott transition coincides with volume 
discontinuities or even changes of the crystal symmetry. This applies, for example, to the 
canonical Mott-Hubbard systems VO$_2$ (Ref.\ \onlinecite{Lupi07}) and vanadium sesquioxide doped with chromium, 
(V$_{0.95}$Cr$_{0.04}$)$_2$O$_3$,\cite{McWhan70,McWhan71,McWhan73} 
and also to MnO,\cite{Yoo05} YNiO$_3$,\cite{Garcia-Munoz03} 
Fe$_2$O$_3$,\cite{Rozenberg02} or FeI$_2$.\cite{Rozenberg03} It was even suggested 
that as a rule the Mott transition coincides with a structural phase transition and volume 
collapse.\cite{Rozenberg03}  Our finding of a pressure-induced structural phase 
transition in TiOBr and TiOCl at the same pressure where the Mott-Hubbard gap closes, 
is in agreement with such an interpretation of the Mott-Hubbard transition.

The importance of electronic correlations for the underlying mechanism of the observed
gap closure in TiO$X$ is suggested by the effective mass of the charge
carriers, as estimated from the spectral weight analysis. 
As demonstrated in Sec. \ref{transmittance}, for both TiOBr and TiOCl the spectral weight becomes 
pressure-independent above a certain pressure [see inset of 
Fig. \ref{fig:cond-all} and Fig. 4(b) in Ref. \onlinecite{Kuntscher06}]. From the high-pressure
value of the spectral weight one can estimate an effective density of charge carriers
according to Eq.\ (\ref{carrierdensity}),\cite{comment1} averaged over
the two studied crystal directions, to $n_{eff}$=$(0.6 \pm 0.2)$$\cdot$$10^{21}$cm$^{-3}$ for TiOBr
and $n_{eff}$=$(1.3 \pm 0.2)$$\cdot$$10^{21}$cm$^{-3}$ for TiOCl for the same frequency range.\cite{Kuntscher06}
Based on these values the effective number of charge carriers per Ti atom, $N_{eff}$, can 
in principle be calculated, if the unit cell volume and the number of formula units per unit 
cell are known.
For an estimate
of $N_{eff}$ we assumed a high-pressure volume of 93 \AA$^3$ (82 \AA$^3$) and a number of formula units
per unit cell of $Z$=2 ($Z$=2) for TiOBr (TiOCl).
Hereby, we neglected the change of the crystal symmetry and a possible collapse of the unit cell volume
at the insulator-to-metal transition; the latter effect usually ranges between 1 and 10 \%.
\cite{McWhan70,Yoo05,Rozenberg02,Rozenberg03}
Under these assumptions we obtained $N_{eff}$=0.03 $\pm$0.01 for TiOBr and $N_{eff}$=0.05 $\pm$0.01
in the case of TiOCl. I. e., $N_{eff}$ is much lower than the expected value of 1.

One possible explanation for the reduced value of $N_{eff}$ could be that the charge carriers 
only partly contribute to the excitations in the specified frequency range. In addition,
the reduction might be related to an enhanced effective mass of the charge carriers, typically found 
in materials with strong electronic correlations. The mass enhancement in TiO$X$ might get stronger when
the system approaches the Mott insulating state, as suggested by the suppressed carrier density
with decreasing pressure [see inset of Fig. \ref{fig:cond-all} and Fig. 4(b) in 
Ref. \onlinecite{Kuntscher06}]. A mass enhancement in the vicinity of a transition to a Mott
insulator was theoretically predicted \cite{Brinkman70} and observed in some cases.
\cite{Qazilbash06,Merino08}

In order to understand the main mechanism driving the observed closure of the Mott-Hubbard gap,
the crystal structure of the high-pressure phase might be an important piece of information. 
However, up to now we could not resolve the symmetry of the crystal structure at high pressures. 
In this regard, density-functional calculations \cite{Valenti08} might provide predictions 
which could then be tested on our x-ray diffraction data.

Finally, we would like to comment on the possibility of the metallic character of the high-pressure
phase in TiO$X$.
Based on our data we cannot prove the existence of a Drude term in the optical response related to 
coherent quasiparticles at high pressures.\cite{Rozenberg95} 
It was, however, demonstrated theoretically and experimentally in various cases, that 
above a certain temperature the absence of a Drude term in a correlated system located on the metallic 
side of the Mott transition is to be expected:
A lot of theoretical work has been devoted to the transport properties of systems close to the
first-order Mott transition at low temperatures and in the crossover regime at elevated temperatures.
Optical conductivity spectra for different interaction strengths and different temperatures 
were obtained in a dynamical mean-field theory (DMFT) treatment of the Hubbard model.\cite{Georges96}
It was shown that only below a certain temperature $T_{coh}$ a quasiparticle peak involving
coherent excitations appears at the Fermi energy and the Fermi liquid description applies. 
As a result, only at low temperatures ($T$$<$$T_{coh}$) a Drude term should be present in the 
optical conductivity spectrum. With increasing temperature, the quasiparticle peak is gradually destroyed
and disappears above the temperature $T_{coh}$. 
Such a behavior was demonstrated for the two-dimensional organic charge-transfer salts 
$\kappa$-(BEDT-TTF)$_2$Cu[N(CN)$_2$]Br$_x$Cl$_{1-x}$:\cite{Merino08} Even for a high Br content, i.e., on the 
metallic side of the Mott transition, no Drude-like peak is present down to approx.\ 50~K.
Only below this temperature a Drude-like feature appears, which can be described with an extended
Drude model, with a frequency-dependent scattering rate and effective mass. 

The optical conductivity spectra of TiO$X$ as a function of pressure were obtained at room temperature. 
According to the findings for organic salts mentioned above and in other cases, \cite{Georges96} the 
seeming absence of a Drude-like contribution in the optical response of TiO$X$ at high pressures could 
be explained by the elevated measurement temperature. Still, the metallic state appears to be the 
most plausible high-pressure phase for TiO$X$ based on our experimental results. The shape of the 
optical conductivity spectra at high pressures is, however, an open issue. Furthermore, a direct
proof of the Drude response might be obtainable by pressure-dependent reflectance measurements 
carried out at low temperatures.

%Based on our data we cannot definitely prove the existence of a Drude term,
%related to a coherent quasiparticle peak at the Fermi energy,\cite{Rozenberg95} in the high-pressure
%phase. Nevertheless, the metallic state appears to be the most plausible phase at high pressures
%based on our experimental results.

\section{Conclusions}
\label{summary}
In conclusion, we have studied the pressure-dependent optical response of TiOBr and TiOCl at room temperature
by transmittance and reflectance measurements in combination with pressure-dependent x-ray powder
diffraction experiments. For both compounds the infrared transmittance is suppressed above a critical pressure. 
The pressure-dependent reflectance and corresponding optical conductivity spectra reveal additional electronic 
excitations at high pressures extending down to the far-infrared range. These findings suggest the closure 
of the Mott-Hubbard gap under pressure. 
For TiOBr the pressure-induced suppression of the infrared transmittance also occurs at 23~K, where the 
compounds is in the spin-Peierls phase at ambient pressure.
The orbital excitations in TiOBr shift linearly to higher frequency with increasing pressure. The shifts
are not completely reversible upon pressure release, especially at low temperatures. 

The pressure-induced changes occur at somewhat lower pressure in the case of TiOBr compared
to TiOCl. This difference cannot be attributed to the more two-dimensional character of TiOBr, since
according to the ratio of the crystal parameters $a$ and $b$ the system becomes more one-dimensional
under pressure, i.e., the high-pressure state seems to be rather of one-dimenisonal than of 
two-dimensional character. 

The closure of the Mott-Hubbard gap coincides with a structural phase transition. 
From the results of our pressure-dependent x-ray powder diffraction measurements on TiOBr and TiOCl 
we could furthermore 
extract the pressure-dependence of the lattice parameters and of the unit cell volume. 
The latter can be well described by the Murnaghan equation. 
The enhancement of the effective mass of the charge carriers around the critical pressure
suggests the importance of electronic correlations for the mechanism driving the transition. 
However, the lattice degree of freedom seems to play in important role as well, since the 
crystal symmetry changes at the transition pressure.

\subsection*{Acknowledgements}
We acknowledge the ANKA Angstr\"omquelle Karlsruhe for the provision
of beamtime and we would like to thank B. Gasharova, Y.-L. Mathis, D. Moss,
and M. S\"upfle for assistance using the beamline ANKA-IR. Facilities and
beamtime provided by the European Synchrotron Radiation Facility is gratefully acknowledged.
We furthermore thank K. Syassen for providing valuable information about the optical design 
of the infrared microscope with large working distance. Fruitful discussions with Jan Kunes 
are greatfully acknowledged.
Financial support by the DFG, including the Emmy Noether-program, SFB 484, and DFG-CL124/6-1, 
is acknowledged.


\begin{thebibliography}{10}
\bibitem{comment1}
For the estimation of the effective carrier density $n_{eff}$ we used the free electron mass. Furthermore,
we assumed a background dielectric constant $\epsilon_\infty$$\approx$3 in the Drude-Lorentz fitting,
as described in Sec. \ref{transmittance}.
\end{thebibliography}
\end{document}